\documentclass[%
 reprint,
 amsmath,amssymb,
 aps,
]{revtex4-1}

\usepackage{graphicx}
\usepackage{dcolumn}
\usepackage{bm}

\begin{document}

\preprint{PUHEP-TH-1001}

\title{Hexad Preons and Emergent Gravity in 3-dimensional Complex
Spacetime}

\author{Shun-Zhi Wang~$^{1,2}$}
\email{wsz08@pku.edu.cn}
 \affiliation{%
 $^{1}$College of Fundamental Studies,\\
 Shanghai University of Engineering Science,
 Shanghai, 201620, P.R.China\\
 $^{2}$Center for High Energy Physics, Peking University,
 Beijing, 100871,  P.R.China
}%

\date{\today}

\begin{abstract}
We suggest that at high energy each space dimension has their
own time dimension, forming a 3-dimensional complex spacetime.
Based on this hypothesis, we propose that the primordial universe is made of six fundamental anticommutative fields and their complex conjugate states. These fields are called Hexad Preons which carry hypercolor degree of freedom transforming under $U(3,3)$ gauge group.
Upon the breakdown of the gauge group $U(3,3)$ to its maximal compact subgroup, $U(3)\otimes U(3)$, the metric emerges . Leptons, quarks, as well as other matter states may be formed from the subsequent condensate of Hexad Preons. Strong and electroweak forces are manifestations of the hypercolor interaction in the corresponding cases. Our framework may shed light on many problems in cosmology and particle physics.

\begin{description}
\item[PACS numbers]
02.40.Tt,
04.50.-h,
 11.10.Kk,
 12.10.-g
\end{description}
\end{abstract}

\pacs{Valid PACS appear here}

\maketitle


\section{\label{sec:intro}Introduction}

It is known that there are four fundamental interactions: gravity, weak, electromagnetic and strong interactions. Gravity is described by the Einstein's general relativity  while the other three interactions are incorporated in the Standard Model (SM) of particle physics. Though general relativity and SM have been confirmed to very high accuracy\cite{PDG}, many problems remain unresolved.
The nature of gravity at small distance is unknown. Current cosmological models based on general relativity must answer such questions like dark matter, dark energy, and baryon asymmetries. In SM, the Quantum Chromodynamics(QCD) and Quantum electroweak theory(GWS) are not genuinely unified. Other questions include: Why gauge group $SU(3)_C\otimes SU(2)_L\otimes U(1)_Y$? Why replication of leptons and quarks? How to interpret the fermion mass spectrum and fermion mixing?

Many attempts, including  extra dimensions\cite{[{For a introduction, see~}]EDs},
supersymmetry\cite{SUSY}, grand unified theories(GUTs)\cite{GUTs}, technicolor [\onlinecite{[{ For a introductions and reviews, see~}] Hill:2002ap}] and compositeness [\onlinecite{[{For a review, see~}] Harari:1984us}],
have been made to resolve these problems. However, the solutions remain obscure.

Spacetime plays the fundamental role in physics.
Because of the great success of general relativity
and the SM, Poincar\'e invariance in four dimensional spacetime
(three space dimensions plus one time dimension)
has been the foundation of modern theoretical physics.
Moreover, Poincar\'e invariance is still the guiding
principle for many new physics beyond the SM, e. g.,
supersymmetry and string theory\cite{string}.

However, we'd better to keep an open mind on the nature of
spacetime at very high energy.
Here we would point out two aspects about the Poincar\'e invariance in four dimensional
spacetime.
The first is about the general coordinate covariance.
In order to consider general relativity in a unified
theoretical framework, we must require general coordinate
covariance rather than Poincar\'e invariance.
The second is the obvious asymmetry between space dimensions and time dimension in the sense that the three space dimensions share the same time dimension. Though at low energy it is sufficient to describe the world in a four dimensional spacetime continuum, it is not necessary so at all energy scales.

The idea of time-like extra dimensions is not new. The early works on this subject are mainly concerned with the conformal aspects of field theory\cite{Dirac1936fq,Kastrup1966zz,Mack1969rr}.
Most time-like extra dimension models are Kaluza-Klein type theory
with compact extra time dimensions \cite{Pavsic1976ta,Ingraham1977pj,Ingraham1977pk,Ingraham1978pn,
Ingraham1978pm,
Ingraham1978pp,Ingraham1978ka,Ingraham1982xe,Ingraham1982xd,
Matsuda2000cz,Matsuda2000nk}.
The constrains from unitarity and causality on the maximum
radius of the internal time-like directions have been investigated
\cite{Dvali:1999hn,Yndurain:1990fq,Erdem:2005wa,Quiros:2007ym}. Gravitation, electrodynamics and electroweak interaction were reconsidered in the six-dimensional spacetime with three times\cite{Lunsford:2004yd, Taylor:1979sd,Dartora:2009zz}.
Other developments include time-like extra dimensions in the brane world scenarios\cite{Gogberashvili:2000yq, Gogberashvili:2000hu, Bonelli:2000tz, Chaichian:2000az, Berezhiani:2001bn, Iglesias:2000iz,
Li:2000kx,Yusofi:2010zz}, F-theory with one time-like extra dimension\cite{Vafa:1996xn}, supergravity and string theory involving a holographic principle and extra time dimensions\cite{Maldacena:1997re,Witten:1998qj,Hull:1999mr},
as well as quantum theory in spacetime with one time-like extra dimension\cite{Koch:2008hn, Foster:2010vt}. In addition, the so called "two-time physics (2T-physics)" where the phase space is gauged, has established the unitarity and causality in 4+2 dimensions with one extra space
and one extra time dimensions and have obtained many interesting results [\onlinecite{[][{, and references therein.}]Bars:2010xi}].

In this paper, we suggest that at high energy the space
dimension remains three and each space dimension has
their own time dimension, forming a 3-dimensional complex
spacetime. We will show that this simple assumption has
many important consequences. In the following we will first
give a mathematical account for complex spacetime within the framework of Generalized Complex Geometry\cite{GCG}
and then,
in a subsequent sections, we will explore the relation between this
mathematical constructure and the emergent fundamental interactions.

\section{\label{sec:GCG}Generalized Complex Spacetime Manifold}

Suppose $M$ is a six dimensional smooth manifold. Let
$\{x^i, y^i, i=1, 2, 3\}$ be the real spacetime coordinates in a chart
$\{U,\phi\}$ around $p \in M$. At every spacetime point $p$ there is a tangent space $T_p(M)$ and a cotangent space $T_p^*(M)$. If every point $p$ is assigned a vector from $T_p(M)$, we have a tangent vector field. The space of tangent vector fields is denoted by $TM$. Similar is for the space of cotangent vector fields, $T^*M$. In terms of the coordinate basis, tangent vector fields are spanned
by $\{\partial /\partial x^i, \partial
/\partial y^i, i=1, 2, 3 \}$ and the cotangent vector fields by $\{dx^i, dy^i, i=1, 2, 3 \}$. $T^*M$ is dual to $TM$ and the coordinate basis are dual to each other:
\begin{eqnarray}
&&dx^i(\frac{\partial}{\partial x^j}) =  dy^i(\frac{\partial}{\partial y^j}) =\delta^i_j,~i,j=1, 2, 3,\nonumber\\ &&dx^i(\frac{\partial}{\partial y^j})=dy^i(\frac{\partial}{\partial x^j})=0,~i,j=1, 2, 3.\nonumber
\end{eqnarray}

Consider the space $TM\oplus T^*M=\{X+\xi, X\in TM,\xi \in T^*M\}$. There is an \emph{inner product} $<,>$ defined by
\begin{equation}
<X+\xi, Y+\eta>=\frac{1}{2}(\xi(Y)+\eta(X)),
\label{natural-pairing}
\end{equation}
where $X, Y \in TM$ and $\xi, \eta \in T^*M$. This inner product is symmetric and non-degenerate.

The linear mappings from $TM\oplus T^*M$ to itself
which preserve the inner product $<,>$ are called orthogonal transformations. All orthogonal transformations form a group $O(TM\oplus T^*M)$ called the orthogonal group. The inner product $<,>$
has signature (6, 6), so we have non-compact orthogonal group $O(TM \oplus T^*M)\cong O(6,6)$.

One important class of orthogonal transformations is $B$- field transformations. $B$ field is a smooth 2-form, a bilinear mapping with the property that for all $X, Y \in TM, B(X, Y)=-B(Y, X)$. $B$ may be viewed as a map $TM\rightarrow T^*M$ in the sense that for every $X\in TM, B(X)(Y)=B(X, Y)$.  In the form of block matrix,  $B$ may be written as
\[B~\sim\left (\begin{array}{cc}
 0&0\\
 B&0\end{array}\right)
\]
Taking the usual matrix exponential yields the $B$- field transformation
\begin{equation}
\textrm{exp}(B)=\left (\begin{array}{cc}
 1&0\\
 B&1\end{array}\right).
\end{equation}
that is, $B$- field transformations send $X\oplus \xi\in TM\oplus T^*M $ to $X\oplus (B(X)+\xi)\in TM\oplus T^*M $. Obviously the $B$- field transformations preserve the inner product $<,>$.

In order to reduce the orthogonal group $O(6, 6)$ to $U(3, 3)$, a generalized complex structure $J$ is introduced on $TM\oplus T^*M$ as follows.
\begin{eqnarray}
J(\frac{\partial}{\partial x^i})=\frac{\partial}{\partial y^i},~&&~ J(\frac{\partial}{\partial y^i})=-\frac{\partial}{\partial x^i},\nonumber\\
J(dx^i)=dy^i,~&&~J(dy^i)=-dx^i.\nonumber
\end{eqnarray}
This is a linear map $J: TM\oplus T^*M \rightarrow TM\oplus T^*M$ which satisfies $J^2=-1$ and preserves the inner product
\[<Jv_1, Jv_2>=<v_1,v_2>,~~v_1,v_2\in TM\oplus T^*M.\]

In terms of above generalized complex structure $J$, the spacetime manifold $M$ can be turned into a complex manifold with coordinates $z^i$: \[z^i=x^i+\sqrt{-1}y^i,~~~~ i=1, 2, 3.\]
In the following we will use the convention that
$\overline{z^i}=\bar{z}_{i}=x^i-\sqrt{-1}y^i, i=1, 2,
3$. We take the coordinate basis of complexified space $(TM\oplus T^*M)\otimes C$ as $\{\partial_i, d\bar{z}_i, dz^i,
\bar{\partial}^i, i=1, 2,3\}$, where
\begin{eqnarray}
\partial_i\equiv \frac{\partial}{\partial z^i}&=&\frac{1}{2}(\frac{\partial}{\partial x^i}-\sqrt{-1}\frac{\partial}{\partial y^i}),\nonumber \\
\bar{\partial}^{i}\equiv \frac{\partial}{\partial \bar{z}_{i}}&=&\frac{1}{2}(\frac{\partial}{\partial x^i}+\sqrt{-1}\frac{\partial}{\partial y^i}),\nonumber\\
dz^i&=&dx^i+\sqrt{-1}dy^i,\nonumber\\
d\bar{z}_i&=&dx^i-\sqrt{-1}dy^i.
\end{eqnarray}
The structure $J$ can be extended to the space $(TM\oplus T^*M)\otimes C$ so that
\begin{eqnarray}
J(\partial_i)=\sqrt{-1}\partial_i,~&& J(d\bar{z}_i)=\sqrt{-1}d\bar{z}_i, \nonumber\\
J(\bar{\partial}^i)=-\sqrt{-1}\bar{\partial}^i,~&&
J(dz^i)=-\sqrt{-1}dz^i.\nonumber
\end{eqnarray}
Therefore we can devide the
space $(TM\oplus T^*M)\otimes C$ into two parts:
$$(TM\oplus T^*M)\otimes C=V_J^+\oplus V_J^-.$$
The $V_J^+$ is spanned
by the basis $\{\partial_i, d\bar{z}_i\}$ , while the $V_J^-$ is spanned
by the basis $\{dz^i,
\bar{\partial}^i\}$.

Extending the inner product $<,>$ to the complex space $(TM\oplus T^*M)\otimes C$, we have
\begin{eqnarray}
<\partial_i,
\partial_j>=<\partial_i,
d\bar{z}_j>=<d\bar{z}_i,
d\bar{z}_j>=0,\nonumber \\
<\bar{\partial}^i,
\bar{\partial}^j>=<\bar{\partial}^i,
dz^j>=<dz^i,
dz^j>=0,\nonumber\\
<\partial_i,
\bar{\partial}^j>=<d\bar{z}_i,
dz^j>=0,\nonumber\\
<\partial_i,
dz^j>=<d\bar{z}_i,
\bar{\partial}^j>=\frac{1}{2}\delta_i^j.
\end{eqnarray}
The next step is to rotate the basis $\{\partial_i, d\bar{z}_i, i=1, 2,3\}$ as follows:
\begin{eqnarray}
(\chi_a, ~\phi_a)&=&(\partial_i, ~d\bar{z}_i)
N^i_a,~a,i=1,2,3\label{inter-states}\\
N&=&\frac{1}{\sqrt{2}}\left (\begin{array}{cc}
 1_{3\times 3}&1_{3\times 3}\\
 1_{3\times 3}&-1_{3\times 3}\end{array}\right).
\label{MHP}
\end{eqnarray}
where $1_{3\times 3}$ stands for $3\times 3$ identity matrix and $N^\dag=N^{-1}=N$. We have
\[<\chi_a,
\bar{\chi}^b>=\frac{1}{2}\delta_a^b,~
<\phi_a,
\bar{\phi}^b>=-\frac{1}{2}\delta_a^b,\]
with other pairings being zero. This shows that the
space $V_J^+$ with the basis $\{\chi_a, \phi_a\}$ has a
inner symmetry $U(3, 3)$, while the space $V_J^-$ with the basis $\{\bar{\chi}^a, \bar{\phi}^a\}$ has a inner symmetry $\overline{U(3, 3)}$.

\section{Hexad Preons and Emergent Forces}

Given the inner product $<,>$, the space $(TM\oplus T^*M)\otimes C$ may be turned into a Clifford algebra by the ralations
\begin{equation}
\{v_1, v_2\}\equiv v_1v_2+v_2v_1=2<v_1,v_2>,\label{Clifford}
\end{equation}
where $v_1, v_2\in (TM\oplus T^*M)\otimes C.$
In particular, we have following anticommutative ralations($i, j=1,2,3$)
\begin{eqnarray}
&\{\partial_i, \partial_j\}=0,~~~\{d\bar{z}_i, d\bar{z}_j\}=0,\label{anti1}\\
&\{\partial_i, d\bar{z}_j\} =0,~~~\label{anti2}
\{\partial_i, dz^{j}\}=\delta_{i}^j,\label{anti3}
 \end{eqnarray}
and their complex conjugate relations.
After a holomorphic coordinate transformation:
\begin{equation}
z^{'i}=z^{'i}(z^1, z^2, z^3),~ i=1,2,3,\label{trans0}
\end{equation}
we have
\begin{eqnarray}
&&\partial'_{i}=\frac{\partial}{\partial z^{'i}}
=\frac{\partial z^j}{\partial z^{'i}}
\frac{\partial}{\partial z^j}
=\frac{\partial z^j}{\partial z^{'i}}\partial_j,~ i=1,2,3,\label{trans1}\\
&&d\bar{z}'_{i}
=\overline{\left(\frac{\partial z^{'i}}{\partial z^{j}}\right)}d\bar{z}_j,~ i=1,2,3.\label{trans2}
 \end{eqnarray}
 Following the physical terminology, we rename the fields $\{\partial_i, \bar{z}_i, i=1, 2, 3\}$ as $\{\psi_i, \bar{\eta}^i, i=1, 2, 3\}$, respectively. Thus we have obtained six
fundamental anticommutative fields $\{\psi_i, \bar{\eta}^i, i=1, 2, 3\}$,
called Hexad Preons.
In analogy with the Dirac spinor, they can be grouped into one multiplet:
\begin{equation}
\Psi=\left (\begin{array}{c}
 \psi^i\\
 \bar{\eta}^i \end{array}\right)~, i=1,2,3.
 \label{HPmass}
\end{equation}
The equations (\ref{trans1})-(\ref{trans2}) show that under the holomorphic coordinate transformation(\ref{trans0}), the Hexad Preons transform in the way similar to that of the components of Dirac spinor:
\begin{eqnarray}
\psi'_{i}=A_i^j\psi_j,~ \bar{\eta}^{'i}=(A^*)^{-1i}_j\bar{\eta}^j,\label{trans3}
 \end{eqnarray}
where $i, j=1,2,3,$ and $A_i^j=\frac{\partial z^j}{\partial z^{'i}}.$

The fields $\{\psi_i, \bar{\eta}^i, i=1, 2, 3\}$ may be called mass eigenstates
of Hexad Preons. After the rotation defined in  Eqn.(\ref{inter-states}),
we obtain the interaction eigenstates $\{\chi_a, \phi_a\}$.
The above discussion shows that $\{\chi_a, \phi_a\}$ carry hypercolor degree of freedom transforming under $U(3,3)$ gauge group.

At this stage there is no metric on the tangent space and no difference between the spacetime coordinates $x^i, y^i, i=1,2,3$. However, by reducing the gauge group $U(3, 3)$ to its maximal compact subgroup $U(3)\otimes U(3)$, we can introduce a generalized metric on $TM\otimes T^*M$ which is compatible with the natural pairing (\ref{natural-pairing}). It is shown\cite{GCG} that this generalized metric is equivalent to a choice of metric $G$ on $TM$ and $B$-field transformation. We may identify this datum $(G, B)$ with a metric $g=g_{ij}dz^i\otimes dz^j+\overline{g_{ij}}d\bar{z}_i\otimes d\bar{z}_j,i,j=1,2,3$ where $g_{ij}$ is given by
\begin{equation}
g_{ij}=G_{ij}+\sqrt{-1}B_{ij}.\nonumber
\end{equation}
In the case of "flat" spacetime where $g_{ij}=\delta_{ij}$, the spacetime interval is
\begin{eqnarray}
ds^2&=&(dz^1)^2+(dz^2)^2+(dz^3)^2+(d\bar{z}_1)^2+(d\bar{z}_2)^2+(d\bar{z}_3)^2\nonumber\\
&=&(dx^1)^2+(dx^2)^2+(dx^3)^2-(dy^1)^2-(dy^2)^2-(dy^3)^2.\nonumber
\end{eqnarray}
 Compared with the spacetime interval in special relativity,   \begin{eqnarray}
ds^2&=&(dx^1)^2+(dx^2)^2+(dx^3)^2-(dt)^2\nonumber,
\end{eqnarray}
we reach the conclusion that at low energy the time we measured is the "length" of the time "vector"
\begin{eqnarray}
dt^2&=&(dy^1)^2+(dy^2)^2+(dy^3)^2.\nonumber\
\end{eqnarray}
This also shows that the difference between the space coordinates $x^i, i=1,2,3,$ and time coordinates $y^i,i=1,2,3,$ is the consequence of symmetry breaking     $U(3, 3)\rightarrow U(3)\otimes U(3)$.

After the symmetry breaking,
the Hexad Preons can also condensate to form subsequent matter states. The symmetry group, $U(3)\otimes U(3)$, means that they can accommodate particle contents of SM easily.

It should be mentioned that our treatment of complex spacetime is intrinsically
different from previous works\cite{woodhouse:1977,Chamseddine:2005at,Chamseddine:2006epa}, where spacetime has four complex dimensions and eventually only the real slice is considered.

\section{Discussion and Conclusions}

 In this paper, it is postulated that at high energy we have three dimensional complex spacetime and at low energy only the "length" of the three time "vector" plus the three space dimensions can be felt. Based on this hypothesis, we can interpret the generalized complex geometry as a mathematical framework to incorporate the four fundamental forces.

The emphasis in this paper is mainly put on the mathematical aspects of the Hexad Preons and emergent gravity in three dimensional complex spacetime. However, we could give some comments on problems mentioned at the beginning of the paper.

 We propose that the primordial universe is made of
six fundamental fields,  Hexad Preons $\left(\psi_i,~\bar{\eta}^i\right)$ in $ \Psi$, and their complex conjugate states
\begin{equation}
\bar{\Psi}=\Psi^\dag\left (\begin{array}{cc}
 0&1_{3\times 3}\\
 1_{3\times 3}&0 \end{array}\right)=\left(\eta^i,~\bar{\psi}_i\right)~, i=1,2,3.
 \label{HPconj}
\end{equation}
In terms of interaction eigenstates, Hexad Preons $\{\chi_a, \phi_a\}$ carry hypercolor degree of freedom
transforming under $U(3,3)$ gauge group. This is the only interaction between the Preons. At this stage
the universe has the biggest internal and spacetime symmetry.

As the gauge group $U(3,3)$
breaks down to its maximal compact subgroup $U(3)\otimes U(3)$, the Hermitian metric emerges so that the gravity may be viewed as a symmetry-breaking effect in quantum field theory. People has been pursued in this direction since 1960's [\onlinecite{[][{, and the references therein.}]Adler:1982ri}] . However, the situation here is more involved because the gauge group is non-compact.

Upon the appearance of metric, complex conjugate operation on
the spacetime coordinates $z^i=x^i+\sqrt{-1}y^i$ means
time reversal. $V_J^-$ contains the complex conjugate states
of Hexad Preons in $V_J^+$, in other words, anti-Preons. If the
anti-Preons in $V_J^-$
condensate forming "vacua", the time reversal symmetry is broken and we have a time arrow. This is closely related to the problems of baryon and CP asymmetries. At the same time, this "negative time" vacua may be the source of (or part of) dark energy.

The subsequent condensate of Hexad Preons, now transforming under $U(3)\otimes U(3)$, may form
leptons, quarks, dark matter, as well as other matter states.
The hypercolor interaction here manifests itself as strong and
electroweak forces in the corresponding cases. Therefore QCD and  GWS theory have the same origin. However, the full recovery of SM dynamics needs the deep understanding of the $U(3)\otimes U(3)$ hypercolor dynamics.  In \cite{Wang:2011ud} we have made a tentative attempt to identify the Preon contents of leptons and quarks. It is shown that all processes in standard model are just reshuffle of Preons. The replication of leptons and quarks can be easily obtained. The gauge group $U(3)\otimes U(3)$ is identified with  $U(1)_Q\otimes SU(3)_C\otimes SU(3)_f\otimes U(1)_w$, which is the smallest extension of the gauge group in Harari-Shupe Model\cite{Harari:1979gi, Shupe:1979fv}. QED and QCD are the same as in the SM while the weak interaction is residual "Van der Waals" forces between preons and dipreons which result in the mixing of fermions.

The dynamics in our Hexad Preons model is different from most of other preon models. Usually it is assumed that the properties of preons are similar to that of QCD [\onlinecite{Eichten:1985fs,*[{See also~}]tHooft1980}]:

(a) The preons are relativistic fermions (Dirac or Majorana) in four dimensional spacetime.

(b) The dypercolor dynamics is described by an unbroken non-abelian gauge symmetry under which preons are nonsinglets.

(c) Leptons, quarks are singlets under the  hypercolor  gauge symmetry.

In our Hexad Preons model, preons are anticommutative fields in the three-dimensional complex spacetime. The hypercolor dynamics is described by a non-compact gauge symmetry. Leptons, quarks and gauge bosons are nonsinglets under the  hypercolor  gauge symmetry. Gravity, weak, electromagnetic and strong interactions may all trace back to the same hypercolor dynamics.

In conclusion, we have shown that in three dimensional complex spacetime, the
 four fundamental forces may have a unique origin. In addition, the picture of universe evolution in this paper is in accord with Y. Nambu's opinion
that \cite{Nambu:2009zz} "As the universe expands and cools down,
it may undergo one or more SSB phase transitions from
states of higher symmetries to lower ones, which change
the governing laws of physics".

\begin{acknowledgments}
This work is supported by Shanghai University of Engineering Science under grant No.2011X34 and by the China Postdoctoral
Science Foundation.
\end{acknowledgments}

\bibliography{apssamp}

\providecommand{\noopsort}[1]{}\providecommand{\singleletter}[1]{#1}%
\begin{thebibliography}{10}%
\makeatletter
\providecommand \@ifxundefined [1]{%
 \ifx #1\undefined \expandafter \@firstoftwo
 \else \expandafter \@secondoftwo
\fi
}%
\providecommand \@ifnum [1]{%
 \ifnum #1\expandafter \@firstoftwo
 \else \expandafter \@secondoftwo
\fi
}%
\providecommand \enquote [1]{``#1''}%
\providecommand \bibnamefont  [1]{#1}%
\providecommand \bibfnamefont [1]{#1}%
\providecommand \citenamefont [1]{#1}%
\providecommand\href[0]{\@sanitize\@href}%
\providecommand\@href[1]{\endgroup\@@startlink{#1}\endgroup\@@href}%
\providecommand\@@href[1]{#1\@@endlink}%
\providecommand \@sanitize [0]{\begingroup\catcode`\&12\catcode`\#12\relax}%
\@ifxundefined \pdfoutput {\@firstoftwo}{%
 \@ifnum{\z@=\pdfoutput}{\@firstoftwo}{\@secondoftwo}%
}{%
 \providecommand\@@startlink[1]{\leavevmode\special{html:<a href="#1">}}%
 \providecommand\@@endlink[0]{\special{html:</a>}}%
}{%
 \providecommand\@@startlink[1]{%
  \leavevmode
  \pdfstartlink
   attr{/Border[0 0 1 ]/H/I/C[0 1 1]}%
   user{/Subtype/Link/A<</Type/Action/S/URI/URI(#1)>>}%
  \relax
 }%
 \providecommand\@@endlink[0]{\pdfendlink}%
}%
\providecommand \url  [0]{\begingroup\@sanitize \@url }%
\providecommand \@url [1]{\endgroup\@href {#1}{\urlprefix}}%
\providecommand \urlprefix [0]{URL }%
\providecommand \Eprint[0]{\href }%
\@ifxundefined \urlstyle {%
  \providecommand \doi [1]{doi:\discretionary{}{}{}#1}%
}{%
  \providecommand \doi [0]{doi:\discretionary{}{}{}\begingroup
  \urlstyle{rm}\Url }%
}%
\providecommand \doibase [0]{http://dx.doi.org/}%
\providecommand \Doi[1]{\href{\doibase#1}}%
\providecommand \bibAnnote [3]{%
  \BibitemShut{#1}%
  \begin{quotation}\noindent
    \textsc{Key:}\ #2\\\textsc{Annotation:}\ #3%
  \end{quotation}%
}%
\providecommand \bibAnnoteFile [2]{%
  \IfFileExists{#2}{\bibAnnote {#1} {#2} {\input{#2}}}{}%
}%
\providecommand \typeout [0]{\immediate \write \m@ne }%
\providecommand \selectlanguage [0]{\@gobble}%
\providecommand \bibinfo [0]{\@secondoftwo}%
\providecommand \bibfield [0]{\@secondoftwo}%
\providecommand \translation [1]{[#1]}%
\providecommand \BibitemOpen[0]{}%
\providecommand \bibitemStop [0]{}%
\providecommand \bibitemNoStop [0]{.\EOS\space}%
\providecommand \EOS [0]{\spacefactor3000\relax}%
\providecommand \BibitemShut [1]{\csname bibitem#1\endcsname}%
\bibitem{PDG}%
  \BibitemOpen
  \bibfield{author}{%
  \bibinfo {author} {\bibfnamefont{C.}~\bibnamefont{Amsler}} \emph{et~al.}
  (\bibinfo {collaboration} {Particle Data Group}),\ }%
  \bibfield{journal}{%
  \bibinfo {journal} {Phys.\ Lett.}\ }%
  \textbf{\bibinfo {volume} {B667}},\ \bibinfo {pages} {1} (\bibinfo {year}
  {2008})%
  \bibAnnoteFile{NoStop}{PDG}%
\bibitem{EDs}%
  \BibitemOpen
  \bibfield{author}{%
  \bibinfo {author} {\bibfnamefont{C.}~\bibnamefont{Csaki}}}%
   (\bibinfo {year} {2004}),\
  \Eprint{http://arxiv.org/abs/hep-ph/0404096}{arXiv:hep-ph/0404096}%
  \bibAnnoteFile{NoStop}{EDs}%
\bibitem{SUSY}%
  \BibitemOpen
  \bibfield{author}{%
  \bibinfo {author} {\bibfnamefont{J.}~\bibnamefont{Wess}}\ and\ \bibinfo
  {author} {\bibfnamefont{J.}~\bibnamefont{Bagger}},\ }%
  \emph{\bibinfo {title} {Supersymmetry and supergravity}}\ (\bibinfo
  {publisher} {Princeton University Press},\ \bibinfo {year} {1992})%
  \bibAnnoteFile{NoStop}{SUSY}%
\bibitem{GUTs}%
  \BibitemOpen
  \bibfield{author}{%
  \bibinfo {author} {\bibfnamefont{G.}~\bibnamefont{Ross}},\ }%
  \emph{\bibinfo {title} {Grand unified theories}}\ (\bibinfo {publisher}
  {Westview Press},\ \bibinfo {year} {1984})%
  \bibAnnoteFile{NoStop}{GUTs}%
\bibitem{Hill:2002ap}%
  \BibitemOpen
  \bibfield{author}{%
  \bibinfo {author} {\bibfnamefont{C.~T.}\ \bibnamefont{Hill}}\ and\ \bibinfo
  {author} {\bibfnamefont{E.~H.}\ \bibnamefont{Simmons}},\ }%
  \bibfield{journal}{%
  \Doi{10.1016/S0370-1573(03)00140-6}{\bibinfo {journal} {Phys. Rept.}}\ }%
  \textbf{\bibinfo {volume} {381}},\ \bibinfo {pages} {235} (\bibinfo {year}
  {2003}),\ \Eprint{http://arxiv.org/abs/hep-ph/0203079}{arXiv:hep-ph/0203079}%
  \bibAnnoteFile{NoStop}{Hill:2002ap}%
\bibitem{Harari:1984us}%
  \BibitemOpen
  \bibfield{author}{%
  \bibinfo {author} {\bibfnamefont{H.}~\bibnamefont{Harari}}\ }%
  \bibinfo {note} {in *St. Andrews 1984, Proceedings, Fundamental Forces*, 357-
  398}%
  \bibAnnoteFile{NoStop}{Harari:1984us}%
\bibitem{string}%
  \BibitemOpen
  \bibfield{author}{%
  \bibinfo {author} {\bibfnamefont{M.}~\bibnamefont{Green}}, \bibinfo {author}
  {\bibfnamefont{J.~H.}\ \bibnamefont{Schwarz}},\ and\ \bibinfo {author}
  {\bibfnamefont{E.}~\bibnamefont{Witten}},\ }%
  \emph{\bibinfo {title} {Superstring theory}}\ (\bibinfo {publisher}
  {Cambridge University Press},\ \bibinfo {year} {1987})%
  \bibAnnoteFile{NoStop}{string}%
\bibitem{Dirac1936fq}%
  \BibitemOpen
  \bibfield{author}{%
  \bibinfo {author} {\bibfnamefont{P.~A.~M.}\ \bibnamefont{Dirac}},\ }%
  \bibfield{journal}{%
  \bibinfo {journal} {Annals Math.}\ }%
  \textbf{\bibinfo {volume} {37}},\ \bibinfo {pages} {429} (\bibinfo {year}
  {1936})%
  \bibAnnoteFile{NoStop}{Dirac1936fq}%
\bibitem{Kastrup1966zz}%
  \BibitemOpen
  \bibfield{author}{%
  \bibinfo {author} {\bibfnamefont{H.~A.}\ \bibnamefont{Kastrup}},\ }%
  \enquote{\bibinfo {title} {{Gauge Properties of the Minkowski Space}},}\
  (\bibinfo {year} {1966})%
  \bibAnnoteFile{NoStop}{Kastrup1966zz}%
\bibitem{Mack1969rr}%
  \BibitemOpen
  \bibfield{author}{%
  \bibinfo {author} {\bibfnamefont{G.}~\bibnamefont{Mack}}\ and\ \bibinfo
  {author} {\bibfnamefont{A.}~\bibnamefont{Salam}},\ }%
  \bibfield{journal}{%
  \Doi{10.1016/0003-4916(69)90278-4}{\bibinfo {journal} {Ann. Phys.}}\ }%
  \textbf{\bibinfo {volume} {53}},\ \bibinfo {pages} {174} (\bibinfo {year}
  {1969})%
  \bibAnnoteFile{NoStop}{Mack1969rr}%
\bibitem{Pavsic1976ta}%
  \BibitemOpen
  \bibfield{author}{%
  \bibinfo {author} {\bibfnamefont{M.}~\bibnamefont{Pavsic}},\ }%
  \bibfield{journal}{%
  \Doi{10.1007/BF02740893}{\bibinfo {journal} {Nuovo Cim.}}\ }%
  \textbf{\bibinfo {volume} {B41}},\ \bibinfo {pages} {397} (\bibinfo {year}
  {1977})%
  \bibAnnoteFile{NoStop}{Pavsic1976ta}%
\bibitem{Ingraham1977pj}%
  \BibitemOpen
  \bibfield{author}{%
  \bibinfo {author} {\bibfnamefont{R.~L.}\ \bibnamefont{Ingraham}},\ }%
  \bibfield{journal}{%
  \Doi{10.1007/BF02748626}{\bibinfo {journal} {Nuovo Cim.}}\ }%
  \textbf{\bibinfo {volume} {B46}},\ \bibinfo {pages} {1} (\bibinfo {year}
  {1978})%
  \bibAnnoteFile{NoStop}{Ingraham1977pj}%
\bibitem{Ingraham1977pk}%
  \BibitemOpen
  \bibfield{author}{%
  \bibinfo {author} {\bibfnamefont{R.~L.}\ \bibnamefont{Ingraham}},\ }%
  \bibfield{journal}{%
  \Doi{10.1007/BF02748627}{\bibinfo {journal} {Nuovo Cim.}}\ }%
  \textbf{\bibinfo {volume} {B46}},\ \bibinfo {pages} {16} (\bibinfo {year}
  {1978})%
  \bibAnnoteFile{NoStop}{Ingraham1977pk}%
\bibitem{Ingraham1978pn}%
  \BibitemOpen
  \bibfield{author}{%
  \bibinfo {author} {\bibfnamefont{R.~L.}\ \bibnamefont{Ingraham}},\ }%
  \bibfield{journal}{%
  \Doi{10.1007/BF02728620}{\bibinfo {journal} {Nuovo Cim.}}\ }%
  \textbf{\bibinfo {volume} {B46}},\ \bibinfo {pages} {217} (\bibinfo {year}
  {1978})%
  \bibAnnoteFile{NoStop}{Ingraham1978pn}%
\bibitem{Ingraham1978pm}%
  \BibitemOpen
  \bibfield{author}{%
  \bibinfo {author} {\bibfnamefont{R.~L.}\ \bibnamefont{Ingraham}},\ }%
  \bibfield{journal}{%
  \Doi{10.1007/BF02728621}{\bibinfo {journal} {Nuovo Cim.}}\ }%
  \textbf{\bibinfo {volume} {B46}},\ \bibinfo {pages} {261} (\bibinfo {year}
  {1978})%
  \bibAnnoteFile{NoStop}{Ingraham1978pm}%
\bibitem{Ingraham1978pp}%
  \BibitemOpen
  \bibfield{author}{%
  \bibinfo {author} {\bibfnamefont{R.~L.}\ \bibnamefont{Ingraham}},\ }%
  \bibfield{journal}{%
  \Doi{10.1007/BF02894587}{\bibinfo {journal} {Nuovo Cim.}}\ }%
  \textbf{\bibinfo {volume} {B47}},\ \bibinfo {pages} {157} (\bibinfo {year}
  {1978})%
  \bibAnnoteFile{NoStop}{Ingraham1978pp}%
\bibitem{Ingraham1978ka}%
  \BibitemOpen
  \bibfield{author}{%
  \bibinfo {author} {\bibfnamefont{R.~L.}\ \bibnamefont{Ingraham}},\ }%
  \bibfield{journal}{%
  \Doi{10.1007/BF02748875}{\bibinfo {journal} {Nuovo Cim.}}\ }%
  \textbf{\bibinfo {volume} {B50}},\ \bibinfo {pages} {233} (\bibinfo {year}
  {1979})%
  \bibAnnoteFile{NoStop}{Ingraham1978ka}%
\bibitem{Ingraham1982xe}%
  \BibitemOpen
  \bibfield{author}{%
  \bibinfo {author} {\bibfnamefont{R.~L.}\ \bibnamefont{Ingraham}},\ }%
  \bibfield{journal}{%
  \Doi{10.1007/BF02890144}{\bibinfo {journal} {Nuovo Cim.}}\ }%
  \textbf{\bibinfo {volume} {B68}},\ \bibinfo {pages} {203} (\bibinfo {year}
  {1982})%
  \bibAnnoteFile{NoStop}{Ingraham1982xe}%
\bibitem{Ingraham1982xd}%
  \BibitemOpen
  \bibfield{author}{%
  \bibinfo {author} {\bibfnamefont{R.~L.}\ \bibnamefont{Ingraham}},\ }%
  \bibfield{journal}{%
  \Doi{10.1007/BF02890145}{\bibinfo {journal} {Nuovo Cim.}}\ }%
  \textbf{\bibinfo {volume} {B68}},\ \bibinfo {pages} {218} (\bibinfo {year}
  {1982})%
  \bibAnnoteFile{NoStop}{Ingraham1982xd}%
\bibitem{Matsuda2000cz}%
  \BibitemOpen
  \bibfield{author}{%
  \bibinfo {author} {\bibfnamefont{S.}~\bibnamefont{Matsuda}}\ and\ \bibinfo
  {author} {\bibfnamefont{S.}~\bibnamefont{Seki}},\ }%
  \bibfield{journal}{%
  \Doi{10.1016/S0550-3213(00)00776-8}{\bibinfo {journal} {Nucl. Phys.}}\ }%
  \textbf{\bibinfo {volume} {B599}},\ \bibinfo {pages} {119} (\bibinfo {year}
  {2001}),\ \Eprint{http://arxiv.org/abs/hep-th/0008216}{arXiv:hep-th/0008216}%
  \bibAnnoteFile{NoStop}{Matsuda2000cz}%
\bibitem{Matsuda2000nk}%
  \BibitemOpen
  \bibfield{author}{%
  \bibinfo {author} {\bibfnamefont{S.}~\bibnamefont{Matsuda}}\ and\ \bibinfo
  {author} {\bibfnamefont{S.}~\bibnamefont{Seki}},\ }%
  \bibfield{journal}{%
  \Doi{10.1103/PhysRevD.63.065014}{\bibinfo {journal} {Phys. Rev.}}\ }%
  \textbf{\bibinfo {volume} {D63}},\ \bibinfo {pages} {065014} (\bibinfo {year}
  {2001}),\ \Eprint{http://arxiv.org/abs/hep-ph/0007290}{arXiv:hep-ph/0007290}%
  \bibAnnoteFile{NoStop}{Matsuda2000nk}%
\bibitem{Dvali:1999hn}%
  \BibitemOpen
  \bibfield{author}{%
  \bibinfo {author} {\bibfnamefont{G.~R.}\ \bibnamefont{Dvali}}, \bibinfo
  {author} {\bibfnamefont{G.}~\bibnamefont{Gabadadze}},\ and\ \bibinfo {author}
  {\bibfnamefont{G.}~\bibnamefont{Senjanovic}}}%
   (\bibinfo {year} {1999}),\
  \Eprint{http://arxiv.org/abs/hep-ph/9910207}{arXiv:hep-ph/9910207}%
  \bibAnnoteFile{NoStop}{Dvali:1999hn}%
\bibitem{Yndurain:1990fq}%
  \BibitemOpen
  \bibfield{author}{%
  \bibinfo {author} {\bibfnamefont{F.~J.}\ \bibnamefont{Yndurain}},\ }%
  \bibfield{journal}{%
  \Doi{10.1016/0370-2693(91)90210-H}{\bibinfo {journal} {Phys. Lett.}}\ }%
  \textbf{\bibinfo {volume} {B256}},\ \bibinfo {pages} {15} (\bibinfo {year}
  {1991})%
  \bibAnnoteFile{NoStop}{Yndurain:1990fq}%
\bibitem{Erdem:2005wa}%
  \BibitemOpen
  \bibfield{author}{%
  \bibinfo {author} {\bibfnamefont{R.}~\bibnamefont{Erdem}}\ and\ \bibinfo
  {author} {\bibfnamefont{C.~S.}\ \bibnamefont{Un}},\ }%
  \bibfield{journal}{%
  \Doi{10.1140/epjc/s2006-02587-5}{\bibinfo {journal} {Eur. Phys. J.}}\ }%
  \textbf{\bibinfo {volume} {C47}},\ \bibinfo {pages} {845} (\bibinfo {year}
  {2006}),\ \Eprint{http://arxiv.org/abs/hep-ph/0510207}{arXiv:hep-ph/0510207}%
  \bibAnnoteFile{NoStop}{Erdem:2005wa}%
\bibitem{Quiros:2007ym}%
  \BibitemOpen
  \bibfield{author}{%
  \bibinfo {author} {\bibfnamefont{I.}~\bibnamefont{Quiros}}}%
   (\bibinfo {year} {2007}),\
  \Eprint{http://arxiv.org/abs/0707.0714}{arXiv:0707.0714 [gr-qc]}%
  \bibAnnoteFile{NoStop}{Quiros:2007ym}%
\bibitem{Lunsford:2004yd}%
  \BibitemOpen
  \bibfield{author}{%
  \bibinfo {author} {\bibfnamefont{D.~R.}\ \bibnamefont{Lunsford}},\ }%
  \bibfield{journal}{%
  \Doi{10.1023/B:IJTP.0000028858.08167.81}{\bibinfo {journal} {Int. J. Theor.
  Phys.}}\ }%
  \textbf{\bibinfo {volume} {43}},\ \bibinfo {pages} {161} (\bibinfo {year}
  {2004})%
  \bibAnnoteFile{NoStop}{Lunsford:2004yd}%
\bibitem{Taylor:1979sd}%
  \BibitemOpen
  \bibfield{author}{%
  \bibinfo {author} {\bibfnamefont{J.~G.}\ \bibnamefont{Taylor}},\ }%
  \bibfield{journal}{%
  \bibinfo {journal} {J. Phys.}\ }%
  \textbf{\bibinfo {volume} {A13}},\ \bibinfo {pages} {1861} (\bibinfo {year}
  {1980})%
  \bibAnnoteFile{NoStop}{Taylor:1979sd}%
\bibitem{Dartora:2009zz}%
  \BibitemOpen
  \bibfield{author}{%
  \bibinfo {author} {\bibfnamefont{C.~A.}\ \bibnamefont{Dartora}}\ and\
  \bibinfo {author} {\bibfnamefont{G.~G.}\ \bibnamefont{Cabrera}},\ }%
  \bibfield{journal}{%
  \Doi{10.1007/s10773-009-0177-9}{\bibinfo {journal} {Int. J. Theor. Phys.}}\
  }%
  \textbf{\bibinfo {volume} {49}},\ \bibinfo {pages} {51} (\bibinfo {year}
  {2010})%
  \bibAnnoteFile{NoStop}{Dartora:2009zz}%
\bibitem{Gogberashvili:2000yq}%
  \BibitemOpen
  \bibfield{author}{%
  \bibinfo {author} {\bibfnamefont{M.}~\bibnamefont{Gogberashvili}},\ }%
  \bibfield{journal}{%
  \Doi{10.1016/S0370-2693(00)00614-6}{\bibinfo {journal} {Phys. Lett.}}\ }%
  \textbf{\bibinfo {volume} {B484}},\ \bibinfo {pages} {124} (\bibinfo {year}
  {2000}),\ \Eprint{http://arxiv.org/abs/hep-ph/0001109}{arXiv:hep-ph/0001109}%
  \bibAnnoteFile{NoStop}{Gogberashvili:2000yq}%
\bibitem{Gogberashvili:2000hu}%
  \BibitemOpen
  \bibfield{author}{%
  \bibinfo {author} {\bibfnamefont{M.}~\bibnamefont{Gogberashvili}}\ and\
  \bibinfo {author} {\bibfnamefont{P.}~\bibnamefont{Midodashvili}},\ }%
  \bibfield{journal}{%
  \Doi{10.1016/S0370-2693(01)00782-1}{\bibinfo {journal} {Phys. Lett.}}\ }%
  \textbf{\bibinfo {volume} {B515}},\ \bibinfo {pages} {447} (\bibinfo {year}
  {2001}),\ \Eprint{http://arxiv.org/abs/hep-ph/0005298}{arXiv:hep-ph/0005298}%
  \bibAnnoteFile{NoStop}{Gogberashvili:2000hu}%
\bibitem{Bonelli:2000tz}%
  \BibitemOpen
  \bibfield{author}{%
  \bibinfo {author} {\bibfnamefont{G.}~\bibnamefont{Bonelli}}\ and\ \bibinfo
  {author} {\bibfnamefont{A.}~\bibnamefont{Boyarsky}},\ }%
  \bibfield{journal}{%
  \Doi{10.1016/S0370-2693(00)00985-0}{\bibinfo {journal} {Phys. Lett.}}\ }%
  \textbf{\bibinfo {volume} {B490}},\ \bibinfo {pages} {147} (\bibinfo {year}
  {2000}),\ \Eprint{http://arxiv.org/abs/hep-th/0004058}{arXiv:hep-th/0004058}%
  \bibAnnoteFile{NoStop}{Bonelli:2000tz}%
\bibitem{Chaichian:2000az}%
  \BibitemOpen
  \bibfield{author}{%
  \bibinfo {author} {\bibfnamefont{M.}~\bibnamefont{Chaichian}}\ and\ \bibinfo
  {author} {\bibfnamefont{A.~B.}\ \bibnamefont{Kobakhidze}},\ }%
  \bibfield{journal}{%
  \Doi{10.1016/S0370-2693(00)00874-1}{\bibinfo {journal} {Phys. Lett.}}\ }%
  \textbf{\bibinfo {volume} {B488}},\ \bibinfo {pages} {117} (\bibinfo {year}
  {2000}),\ \Eprint{http://arxiv.org/abs/hep-th/0003269}{arXiv:hep-th/0003269}%
  \bibAnnoteFile{NoStop}{Chaichian:2000az}%
\bibitem{Berezhiani:2001bn}%
  \BibitemOpen
  \bibfield{author}{%
  \bibinfo {author} {\bibfnamefont{Z.}~\bibnamefont{Berezhiani}}, \bibinfo
  {author} {\bibfnamefont{M.}~\bibnamefont{Chaichian}}, \bibinfo {author}
  {\bibfnamefont{A.~B.}\ \bibnamefont{Kobakhidze}},\ and\ \bibinfo {author}
  {\bibfnamefont{Z.~H.}\ \bibnamefont{Yu}},\ }%
  \bibfield{journal}{%
  \Doi{10.1016/S0370-2693(01)01022-X}{\bibinfo {journal} {Phys. Lett.}}\ }%
  \textbf{\bibinfo {volume} {B517}},\ \bibinfo {pages} {387} (\bibinfo {year}
  {2001}),\ \Eprint{http://arxiv.org/abs/hep-th/0102207}{arXiv:hep-th/0102207}%
  \bibAnnoteFile{NoStop}{Berezhiani:2001bn}%
\bibitem{Iglesias:2000iz}%
  \BibitemOpen
  \bibfield{author}{%
  \bibinfo {author} {\bibfnamefont{A.}~\bibnamefont{Iglesias}}\ and\ \bibinfo
  {author} {\bibfnamefont{Z.}~\bibnamefont{Kakushadze}},\ }%
  \bibfield{journal}{%
  \Doi{10.1016/S0370-2693(01)00884-X}{\bibinfo {journal} {Phys. Lett.}}\ }%
  \textbf{\bibinfo {volume} {B515}},\ \bibinfo {pages} {477} (\bibinfo {year}
  {2001}),\ \Eprint{http://arxiv.org/abs/hep-th/0012049}{arXiv:hep-th/0012049}%
  \bibAnnoteFile{NoStop}{Iglesias:2000iz}%
\bibitem{Li:2000kx}%
  \BibitemOpen
  \bibfield{author}{%
  \bibinfo {author} {\bibfnamefont{T.-j.}\ \bibnamefont{Li}},\ }%
  \bibfield{journal}{%
  \Doi{10.1016/S0370-2693(01)00215-5}{\bibinfo {journal} {Phys. Lett.}}\ }%
  \textbf{\bibinfo {volume} {B503}},\ \bibinfo {pages} {163} (\bibinfo {year}
  {2001}),\ \Eprint{http://arxiv.org/abs/hep-th/0009132}{arXiv:hep-th/0009132}%
  \bibAnnoteFile{NoStop}{Li:2000kx}%
\bibitem{Yusofi:2010zz}%
  \BibitemOpen
  \bibfield{author}{%
  \bibinfo {author} {\bibfnamefont{E.}~\bibnamefont{Yusofi}}\ and\ \bibinfo
  {author} {\bibfnamefont{M.}~\bibnamefont{Mohsenzadeh}},\ }%
  \bibfield{journal}{%
  \bibinfo {journal} {Int. J. Theor. Phys.}}%
   (\bibinfo {year} {2010}),\ \doi{\bibinfo {doi} {10.1007/s10773-010-0336-z}}%
  \bibAnnoteFile{NoStop}{Yusofi:2010zz}%
\bibitem{Vafa:1996xn}%
  \BibitemOpen
  \bibfield{author}{%
  \bibinfo {author} {\bibfnamefont{C.}~\bibnamefont{Vafa}},\ }%
  \bibfield{journal}{%
  \Doi{10.1016/0550-3213(96)00172-1}{\bibinfo {journal} {Nucl. Phys.}}\ }%
  \textbf{\bibinfo {volume} {B469}},\ \bibinfo {pages} {403} (\bibinfo {year}
  {1996}),\ \Eprint{http://arxiv.org/abs/hep-th/9602022}{arXiv:hep-th/9602022}%
  \bibAnnoteFile{NoStop}{Vafa:1996xn}%
\bibitem{Maldacena:1997re}%
  \BibitemOpen
  \bibfield{author}{%
  \bibinfo {author} {\bibfnamefont{J.~M.}\ \bibnamefont{Maldacena}},\ }%
  \bibfield{journal}{%
  \bibinfo {journal} {Adv. Theor. Math. Phys.}\ }%
  \textbf{\bibinfo {volume} {2}},\ \bibinfo {pages} {231} (\bibinfo {year}
  {1998}),\ \Eprint{http://arxiv.org/abs/hep-th/9711200}{arXiv:hep-th/9711200}%
  \bibAnnoteFile{NoStop}{Maldacena:1997re}%
\bibitem{Witten:1998qj}%
  \BibitemOpen
  \bibfield{author}{%
  \bibinfo {author} {\bibfnamefont{E.}~\bibnamefont{Witten}},\ }%
  \bibfield{journal}{%
  \bibinfo {journal} {Adv. Theor. Math. Phys.}\ }%
  \textbf{\bibinfo {volume} {2}},\ \bibinfo {pages} {253} (\bibinfo {year}
  {1998}),\ \Eprint{http://arxiv.org/abs/hep-th/9802150}{arXiv:hep-th/9802150}%
  \bibAnnoteFile{NoStop}{Witten:1998qj}%
\bibitem{Hull:1999mr}%
  \BibitemOpen
  \bibfield{author}{%
  \bibinfo {author} {\bibfnamefont{C.~M.}\ \bibnamefont{Hull}}}%
   (\bibinfo {year} {1999}),\
  \Eprint{http://arxiv.org/abs/hep-th/9911080}{arXiv:hep-th/9911080}%
  \bibAnnoteFile{NoStop}{Hull:1999mr}%
\bibitem{Koch:2008hn}%
  \BibitemOpen
  \bibfield{author}{%
  \bibinfo {author} {\bibfnamefont{B.}~\bibnamefont{Koch}}}%
   (\bibinfo {year} {2008}),\
  \Eprint{http://arxiv.org/abs/0801.4635}{arXiv:0801.4635 [quant-ph]}%
  \bibAnnoteFile{NoStop}{Koch:2008hn}%
\bibitem{Foster:2010vt}%
  \BibitemOpen
  \bibfield{author}{%
  \bibinfo {author} {\bibfnamefont{J.~G.}\ \bibnamefont{Foster}}\ and\ \bibinfo
  {author} {\bibfnamefont{B.}~\bibnamefont{Muller}}}%
   (\bibinfo {year} {2010}),\
  \Eprint{http://arxiv.org/abs/1001.2485}{arXiv:1001.2485 [hep-th]}%
  \bibAnnoteFile{NoStop}{Foster:2010vt}%
\bibitem{Bars:2010xi}%
  \BibitemOpen
  \bibfield{author}{%
  \bibinfo {author} {\bibfnamefont{I.}~\bibnamefont{Bars}}}%
   (\bibinfo {year} {2010}),\
  \Eprint{http://arxiv.org/abs/1004.0688}{arXiv:1004.0688 [hep-th]}%
  \bibAnnoteFile{NoStop}{Bars:2010xi}%
\bibitem{GCG}%
  \BibitemOpen
  \bibfield{author}{%
  \bibinfo {author} {\bibfnamefont{M.}~\bibnamefont{Gualtieri}}}%
   (\bibinfo {year} {2003}),\
  \Eprint{http://arxiv.org/abs/math/0401221}{arXiv:math/0401221}%
  \bibAnnoteFile{NoStop}{GCG}%
\bibitem{woodhouse:1977}%
  \BibitemOpen
  \bibfield{author}{%
  \bibinfo {author} {\bibfnamefont{N.}~\bibnamefont{Woodhouse}},\ }%
  \bibfield{journal}{%
  \bibinfo {journal} {Int. J. Theor. Phys.}\ }%
  \textbf{\bibinfo {volume} {16}},\ \bibinfo {pages} {663} (\bibinfo {year}
  {1977})%
  \bibAnnoteFile{NoStop}{woodhouse:1977}%
\bibitem{Chamseddine:2005at}%
  \BibitemOpen
  \bibfield{author}{%
  \bibinfo {author} {\bibfnamefont{A.~H.}\ \bibnamefont{Chamseddine}},\ }%
  \bibfield{journal}{%
  \Doi{10.1007/s00220-005-1466-7}{\bibinfo {journal} {Commun. Math. Phys.}}\ }%
  \textbf{\bibinfo {volume} {264}},\ \bibinfo {pages} {291} (\bibinfo {year}
  {2006}),\ \Eprint{http://arxiv.org/abs/hep-th/0503048}{arXiv:hep-th/0503048}%
  \bibAnnoteFile{NoStop}{Chamseddine:2005at}%
\bibitem{Chamseddine:2006epa}%
  \BibitemOpen
  \bibfield{author}{%
  \bibinfo {author} {\bibfnamefont{A.~H.}\ \bibnamefont{Chamseddine}}}%
   (\bibinfo {year} {2006}),\
  \Eprint{http://arxiv.org/abs/hep-th/0610099}{arXiv:hep-th/0610099}%
  \bibAnnoteFile{NoStop}{Chamseddine:2006epa}%
\bibitem{Adler:1982ri}%
  \BibitemOpen
  \bibfield{author}{%
  \bibinfo {author} {\bibfnamefont{S.~L.}\ \bibnamefont{Adler}},\ }%
  \bibfield{journal}{%
  \Doi{10.1103/RevModPhys.54.729}{\bibinfo {journal} {Rev. Mod. Phys.}}\ }%
  \textbf{\bibinfo {volume} {54}},\ \bibinfo {pages} {729} (\bibinfo {year}
  {1982})%
  \bibAnnoteFile{NoStop}{Adler:1982ri}%
\bibitem{Wang:2011ud}%
  \BibitemOpen
  \bibfield{author}{%
  \bibinfo {author} {\bibfnamefont{S.-Z.}\ \bibnamefont{Wang}}}%
   (\bibinfo {year} {2011}),\
  \Eprint{http://arxiv.org/abs/1112.0181}{arXiv:1112.0181 [hep-ph]}%
  \bibAnnoteFile{NoStop}{Wang:2011ud}%
\bibitem{Harari:1979gi}%
  \BibitemOpen
  \bibfield{author}{%
  \bibinfo {author} {\bibfnamefont{H.}~\bibnamefont{Harari}},\ }%
  \bibfield{journal}{%
  \Doi{10.1016/0370-2693(79)90626-9}{\bibinfo {journal} {Phys. Lett.}}\ }%
  \textbf{\bibinfo {volume} {B86}},\ \bibinfo {pages} {83} (\bibinfo {year}
  {1979})%
  \bibAnnoteFile{NoStop}{Harari:1979gi}%
\bibitem{Shupe:1979fv}%
  \BibitemOpen
  \bibfield{author}{%
  \bibinfo {author} {\bibfnamefont{M.~A.}\ \bibnamefont{Shupe}},\ }%
  \bibfield{journal}{%
  \Doi{10.1016/0370-2693(79)90627-0}{\bibinfo {journal} {Phys. Lett.}}\ }%
  \textbf{\bibinfo {volume} {B86}},\ \bibinfo {pages} {87} (\bibinfo {year}
  {1979})%
  \bibAnnoteFile{NoStop}{Shupe:1979fv}%
\bibitem{Eichten:1985fs}%
  \BibitemOpen
  \bibfield{author}{%
  \bibinfo {author} {\bibfnamefont{E.}~\bibnamefont{Eichten}}, \bibinfo
  {author} {\bibfnamefont{R.~D.}\ \bibnamefont{Peccei}}, \bibinfo {author}
  {\bibfnamefont{J.}~\bibnamefont{Preskill}},\ and\ \bibinfo {author}
  {\bibfnamefont{D.}~\bibnamefont{Zeppenfeld}},\ }%
  \bibfield{journal}{%
  \Doi{10.1016/0550-3213(86)90206-3}{\bibinfo {journal} {Nucl. Phys.}}\ }%
  \textbf{\bibinfo {volume} {B268}},\ \bibinfo {pages} {161} (\bibinfo {year}
  {1986})%
  \bibAnnoteFile{NoStop}{Eichten:1985fs}%
\bibitem{tHooft1980}%
  \BibitemOpen
  \bibfield{author}{%
  \bibinfo {author} {\bibfnamefont{G.}~\bibnamefont{'t~Hooft}},\ }%
  in\ \emph{\bibinfo {booktitle} {Recent Developments in Gauge Theories}},\
  \bibinfo {editor} {edited by\ \bibinfo {editor}
  {\bibfnamefont{G.}~\bibnamefont{'t~Hooft}}}\ (\bibinfo {publisher} {Plenum},\
  \bibinfo {address} {New York},\ \bibinfo {year} {1980})\ p.\ \bibinfo {pages}
  {135}%
  \bibAnnoteFile{NoStop}{tHooft1980}%
\bibitem{Nambu:2009zz}%
  \BibitemOpen
  \bibfield{author}{%
  \bibinfo {author} {\bibfnamefont{Y.}~\bibnamefont{Nambu}},\ }%
  \bibfield{journal}{%
  \bibinfo {journal} {Int.\ J.\ Mod.\ Phys.}\ }%
  \textbf{\bibinfo {volume} {47}},\ \bibinfo {pages} {2371} (\bibinfo {year}
  {2009})%
  \bibAnnoteFile{NoStop}{Nambu:2009zz}%
\end{thebibliography}%

\end{document}